\documentclass[symmetry,article,accept,pdftex,oneauthor]{Definitions/mdpi} 
\firstpage{1} 
\pubvolume{0}
\issuenum{0}
\articlenumber{0}
\pubyear{2025}
\copyrightyear{2025}
\externaleditor{Academic Editor: ...}
\datereceived{...} 
\dateaccepted{...} 
\datepublished{...} 

\hreflink{https://\linebreak 
	doi.org/...} 
\doinum{...}
\pdfoutput=1 

\usepackage{bm}
\usepackage{amsmath,amssymb}
\usepackage{wasysym}
\usepackage{pifont,bbm}
\usepackage{stmaryrd}
\usepackage{graphbox,graphicx}
\newcommand{\xxx}{y_1}
\newcommand{\yyy}{y_2} 
\newcommand{\ktilde}{\tilde\kappa}
\newcommand{\F}{F}

\newcommand{\etamine}{H}

\newcommand{\apj}{Astroph. J.}

\newcommand{\mnras}{Mon. Not. R. Astron. Soc.}
\newcommand{\aap}{Astron. Astrophys.}
\newcommand{\bmath}{\bm}
\newcommand{\be}{\begin{equation}}
\newcommand{\ee}{\end{equation}}

\makeatletter
\renewcommand{\maketag@@@}[1]{\hbox{\m@th\normalsize\normalfont#1}}
\makeatother

\Title{Classical waves and instabilities using the minimalist approach}

\TitleCitation{Classical waves and instabilities using the minimalist approach}

\Author{Nektarios Vlahakis \orcidA{}}

\AuthorNames{Nektarios Vlahakis}

\AuthorCitation{Vlahakis, N.}

\address[1]{%
Department of Physics, National and Kapodistrian University of Athens, University Campus,\linebreak Zografos GR-157 84 Athens, Greece; vlahakis@phys.uoa.gr
} 

\abstract{ 
The minimalist approach in the study of perturbations in fluid dynamics and magnetohydrodynamics involves describing their evolution in the linear regime using a single first-order ordinary differential equation, dubbed \emph{principal equation}. The dispersion relation is determined by requiring that the solution of the principal equation be continuous and satisfy specific boundary conditions for each problem.
The formalism is presented for flows in cartesian geometry and applied to classical cases such as the magnetosonic and gravity waves, the Rayleigh-Taylor instability, and the Kelvin-Helmholtz instability. 
For the latter, we discuss the influence of compressibility and the magnetic field, and also derive analytical expressions for the growth rates and the range of instability in the case of two fluids with the same characteristics. 
}


\begin{document}

\section{Introduction}\label{introduction}

The study of perturbations in any physical system sheds light on the mechanisms responsible for maintaining equilibrium and offers a way to predict temporal behavior in response to variations. Although this task is generally difficult, simplifying each case and focusing on the most critical ingredients determining its evolution is essential. Classical waves and instabilities represent such simplified structures. 
For example, acoustic compressible perturbations, attributed to sound or magnetosonic waves, can be identified even in systems with complex dynamics. Alfv\'en waves play a significant role in magnetized plasmas. When gravity or buoyancy forces are important, differences in density drive gravity waves. By ``gravity'' we may also refer to fictitious forces in non-inertial frames, such as the effective gravity in a decelerating system or centrifugal forces. In cases of heavy over light fluid stratifications, the Rayleigh-Taylor instability comes to mind. Similarly, a relative velocity between two fluids triggers the Kelvin-Helmholtz instability.
These are only a few examples of simplified structures that are crucial for understanding more complicated and realistic settings. They are often connected to temporal behaviors in laboratory experiments or specific features observed in astrophysical systems.

The full study of perturbation evolution in most cases can only be done through numerical simulations. Linearizing the system of equations is a simpler way to understand the conditions under which the system is unstable, although it does not cover nonlinear evolution. Normal mode analysis is a further simplification, allowed only in symmetric unperturbed states, and can be used to find wave frequencies or instability growth rates. This approach allows for simpler parametric studies or even the derivation of analytic expressions that directly reveal the physics of the mechanisms involved.

Since we are discussing classical problems, there are many books, e.g., \cite{1961hhs..book.....C,BS69,1978mit..book.....B,2014idmh.book.....F,Goedbook2}, that provide the most basic information. There are also a plethora of related works in the literature over the years, e.g., \cite{FTZ78,1988ApJ...334...70H,2008A&A...490..493O,2017MNRAS.472.1421M,2019MNRAS.483.2325P,2018MNRAS.475L.125G} to name but a few. Nevertheless, these problems continue to be an active area of research, e.g. \cite{2023MNRAS.524...90C,2024A&A...688A..80K,2024GeoRL..5110477N}, and there is always room -- and need -- to improve our understanding. 

The goal of this paper is to present the methodology to determine the dispersion relation in each case, following the \emph{minimalist approach} introduced in Ref.~\cite{2024Univ...10..183V}. This novel approach simplifies the procedure as much as possible, by using a single first-order ordinary differential equation, dubbed \emph{principal equation}. 
Firstly, we derive this equation through linearization and discuss the necessary boundary conditions (Section~\ref{seclinearanalysis}). Next, we apply the method to determine the dispersion relation for magnetohydrodynamic (MHD) waves (Section~\ref{secmhdwaves}),
gravity waves and the Rayleigh-Taylor instability (Section~\ref{secRayleigh-Taylor}), gravito-acoustic waves (Section~\ref{secgravitoacousticwaves}), and the Kelvin-Helmholtz instability (Section~\ref{seckelvinhelmholtz}). 
We provide a more extensive analysis of the Kelvin-Helmholtz instability, deriving analytical expressions for the growth rate in the magnetized case and discussing the factors that most significantly affect the results.

\section{Linear analysis, the principal equation, and boundary conditions}\label{seclinearanalysis}

Fluid dynamics is governed by the conservation laws of mass, momentum and energy. To include cases of a conducting magnetized fluid (plasma), we combine these laws with Maxwell's equations to obtain the system of  magnetohydrodynamic equations.  
In the non-relativistic limit, and neglecting non-ideal effects such as viscosity, resistivity, and surface tension while using Lorentz-Heaviside units, these equations are:
\begin{eqnarray}
	\dfrac{\partial\rho}{\partial t}
	+\nabla \cdot \left(\rho\bmath V \right) =0 \,, \label{mass}
	\\
	\left( \dfrac{\partial}{\partial t} + {\bmath{V}} \cdot
	\nabla \right)P=c_s^2\left( \dfrac{\partial}{\partial t} + {\bmath{V}} \cdot
	\nabla \right)\rho \,, \label{energy}
	\\
	\rho
	\left( \dfrac{\partial}{\partial t} + {\bmath{V}} \cdot
	\nabla \right)
	{\bmath{V}}=
	-\nabla P +\left(\nabla \times {\bmath{B}}\right) \times \bmath B
	+\rho\bm g \,, \label{momentum}
	\label{mom1}
	\\
	\dfrac{\partial {\bmath{B}}}{\partial t}=\nabla \times \left({\bmath V} \times \bmath B\right)  \,, \label{induction}
	\\ 
	\nabla \cdot \bmath B=0   \,. 
\end{eqnarray}  
We are interested in exploring the perturbations of a steady-state that depends only on one spatial cartesian coordinate. In particular we assume that the unperturbed state has density $\rho_0(x)$, pressure $P(x)$, bulk velocity $\bm V_0=V_{0z}(x)\hat z + V_{0y}(x)\hat y$, magnetic field
$\bm B_0=B_{0z}(x)\hat z + B_{0y}(x)\hat y$, and the acceleration of gravity is $\bm g=-g(x)\hat x$.
Introducing the total pressure 
\begin{eqnarray} 
	\Pi = P + \dfrac{B^2}{2} \,,
	\label{Pi1}  
\end{eqnarray} 
the zeroth order equations are satisfied provided that 
\begin{eqnarray}
	\Pi_0' =-\rho_0g \,.
\end{eqnarray} 
Adding perturbations of the form 
\begin{eqnarray}
	\bmath V =\bm V_0 
	+\left[V_{1z}(x)\hat z+ V_{1x}(x)\hat x + V_{1y}(x)\hat y\right] 
	e^{i(k_yy+k_z z-\omega t)} \,, 
	\\
	\bmath B =\bm B_0
	+\left[B_{1z}(x)\hat z+ B_{1x}(x)\hat x + B_{1y}(x)\hat y\right]e^{i(k_yy+k_z z-\omega t)}\,, 
	\\
	\rho =\rho_0(x)
	+ \rho_1(x)e^{i(k_yy+k_z z-\omega t)} \,,
	\\
	\Pi =\Pi_0(x)
	+ \Pi_1(x)e^{i(k_yy+k_z z-\omega t)}\,,
\end{eqnarray}  
defining the wavevector in the $yz$ plane 
\begin{eqnarray}\bm k_0=k_y\hat y+k_z\hat z\,, \end{eqnarray}
and the Doppler-shifted frequency 
\begin{eqnarray}\omega_0
=\omega-\bm k_0\cdot \bm V_0\,, \end{eqnarray}
we linearize the equations as shown in Appendix~\ref{appendixA}; the main steps and related comments follow. 

The linearized system reduces to two differential equations for the perturbations of the velocity in the $\hat x$ direction and the total pressure (there are algebraic relations for all the other perturbations, connecting them to $V_{1x}$, $\Pi_1$, and their derivatives).  
Instead of $V_{1x}$ and $\Pi_1$ it is more convenient to use two other quantities. 

The first, replacing $V_{1x}$, is the Lagrangian displacement in the $\hat x$ direction.
The relation between the Lagrangian displacement $\bm\xi$ and the velocity perturbation results
from the expression of the velocity in the perturbed location of each fluid element $\bm V+(\bm \xi\cdot\nabla)\bm V\approx \bm V_0+\delta\bm V+(\bm \xi\cdot\nabla)\bm V_0$ which equals $\bm V_0+\dfrac{d\bm\xi}{dt}\approx \bm V_0+\dfrac{\partial\bm\xi}{\partial t}+(\bm V_0\cdot\nabla)\bm \xi$, yielding $\delta\bm V=\dfrac{\partial\bm\xi}{\partial t}+(\bm V_0\cdot\nabla)\bm \xi-(\bm \xi\cdot\nabla)\bm V_0
=-i\omega_0\bm\xi-\xi_x\dfrac{d\bm V_0}{dx}$.
The $\hat x$ component gives $\xi_x=\xxx(x) e^{i(k_yy+k_z z-\omega t)}$ with $V_{1x}=-i\omega_0\xxx$.

The second quantity, replacing $\Pi_1$, is the perturbation of the total pressure in the perturbed location of each fluid element $\yyy=\Pi_1+\xxx \Pi_0'=\Pi_1-\rho_0 g\xxx $. 

The advantage of these replacements is that the new functions $\xxx$ and $\yyy$ are everywhere continuous, even at locations where the unperturbed state has contact discontinuities.

\subsection{The system for $\xxx$, $\yyy$} 
The resulting $2\times 2$ system is
\begin{eqnarray}
\frac{d}{dx} \left( \begin{array}{c}
	y_1 \\ y_2
\end{array} \right) +\left( \begin{array}{cc}
	f_{11} & f_{12} \\ f_{21} & -f_{11}
\end{array} \right) \left( \begin{array}{c}
	y_1 \\ y_2
\end{array} \right)
=0\,, \label{eqsystemxxxyyy}
\\
f_{11}=\dfrac{\rho_0 g}{A} \left(\dfrac{\rho_0\omega_0^2\F^2}{S}-\bm k_0^2\right) \,, 
\quad f_{12}= \dfrac{\ktilde^2}{A} \,,  
\quad f_{21}=-A-\dfrac{\rho_0^2g^2}{A}\left(\dfrac{\F^4}{S}-\bm k_0^2\right)+\rho_0 g'\,, 
\\
\F =\bm k_0\cdot\bm B_0 \,, \quad 
A=\rho_0\omega_0^2-\F^2 \,, \quad
S=\rho_0(Ac_s^2+\omega_0^2B_0^2) \,, \quad
\ktilde^2=\dfrac{\rho_0^2\omega_0^4 }{S}-\bm k_0^2 \,. 
\end{eqnarray}

To simplify the expressions we define the following quantities that have important physical meaning and effect in the resulting dispersion relations through their appearance in the array elements $f_{ij}$. 

The $\F =\bm k_0\cdot \bm B_0$ is connected to the angle between the wavevector and the unperturbed magnetic field. Its presence in a dispersion relation represents the influence of the magnetic tension which acts like a spring with restoring force per mass $-\dfrac{F^2}{\rho_0}\xxx$.
This is evident in the pure Alfv\'en waves (in a static homogeneous plasma without gravity) for which 
the displacement satisfies 
$\ddot y_1=-\omega^2\xxx$ with $\omega=\dfrac{\bm k_0\cdot\bm B_0}{\sqrt{\rho_0}}=\dfrac{F}{\sqrt{\rho_0}}$.

The $A=\rho_0\left(\omega_0^2-\dfrac{\F^2}{\rho_0}\right)$ appears in the denominator of the resulting system of differential equations, and its zeros correspond to the Alfv\'en waves.

The $S=\rho_0^2\left[\left(\dfrac{B_0^2}{\rho_0}+c_s^2\right)\omega_0^2-\dfrac{\F^2}{\rho_0}c_s^2\right]$
is related to the density perturbation in the perturbed location of each fluid element $ \rho_1 +\xxx \rho_0' =
\rho_0^2\dfrac{\omega_0^2\yyy+g\F^2\xxx}{S}$. It is interesting to note that the incompressible limit corresponds to $S\to \infty$.

The $\ktilde$ defined through $\ktilde^2=\dfrac{\rho_0^2\omega_0^4}{S}-\bm k_0^2$ represents the local wavenumber in the $\hat x$ direction. 
Indeed, the latter equation is equivalent to the dispersion relation of the slow/fast magnetosonic waves
$\left(\dfrac{\omega_0^2}{\bm k_t^2}\right)^2
-\left(\dfrac{\omega_0^2}{\bm k_t^2}\right)\left(\dfrac{B_0^2}{\rho_0}+c_s^2\right)+\dfrac{B_0^2}{\rho_0}c_s^2\cos^2\theta_0=0$
if we define the total wavevector $\bm k_t=\bm k_0+\ktilde\hat x$ and the angle $\theta_0$ between $\bm k_t$ and $\bm B_0$ through $\cos\theta_0=\dfrac{\F}{|\bm k_t|B_0}$.
\\ Another way to see the connection between $\ktilde$ and the wavenumber in the $\hat x$ direction is to consider the homogeneous fluid case without gravity. In this scenario, the system~(\ref{eqsystemxxxyyy}) with constant coefficients has solutions $\dfrac{\xxx'}{\xxx}=\dfrac{\yyy'}{\yyy}
=i\ktilde 
$. The equation $\ktilde^2=\dfrac{\rho_0^2\omega_0^4}{S}-\bm k_0^2$ gives two opposite solutions for $ \ktilde$ corresponding to oppositely moving waves in the $\pm \hat x$ direction.
Note that, in general, $\ktilde$ is complex, and its imaginary part corresponds to exponential variation of the eigenfunctions, since $e^{i\ktilde x}=e^{-\Im \ktilde x}e^{i\Re \ktilde x}$. In the incompressible limit, where $S\to\infty$ and $\ktilde^2\to -\bm k_0^2$, it simplifies to purely exponential dependence $e^{-|k_0 x|}$, without sinusoidal dependence on $x$.
More details on these waves will be discussed later; they offer a way to understand the physics of the solutions even in non-homogeneous fluids  and are directly related to the boundary conditions when the fluid extends up to large distances $x\to+\infty$ or $-\infty$.  

The cases $A=0$, $S=0$, $\omega_0=0$, need to be studied separately, something that can be easily done since they lead to simplified equations. These cases correspond to waves and are not necessary for instability studies in which $\bm k_0$ is real and $\omega$ complex. 

\subsection{The principal equation} 

Since the system~(\ref{eqsystemxxxyyy}) is linear only the ratio $Y=\dfrac{\xxx}{\yyy}$ is uniquely defined. 
Using the equations of the system we directly find that $Y$ satisfies the principal equation  
\begin{eqnarray}
	Y' =f_{21} Y^2-2f_{11}Y-f_{12} 
	\,. \label{eqprincipal} \end{eqnarray}
Following the minimalist approach \cite{2024Univ...10..183V} it is sufficient to work with the ratio $Y$ and solve the principal equation in order to find the dispersion relation of the wave or instability.  
We just need to integrate this single, first order differential equation, requiring $Y$ to be everywhere continuous and satisfying the correct boundary conditions at the extreme values of $x$.

Knowing $Y$ we can find all the other perturbations from 
\begin{eqnarray} 
	\dfrac{\yyy'}{\yyy}=-f_{21}Y+f_{11} \,, \quad \xxx=Y\yyy 
	\label{eqsforxxxyyyY}
\end{eqnarray}
and the relations~(\ref{vx5})--(\ref{pressure5}) given in Appendix~\ref{appendixA}. We emphasize though that these are not needed to find the dispersion relation.

\subsection{Boundary conditions}\label{secboundary}

The linearization was based on the functions $\xxx$ and $\yyy$, which are everywhere continuous. The same holds for their ratio. Therefore, in the case of discontinuities in the unperturbed state, we simply continue the integration of the principal equation when passing from one medium to another, keeping $Y$ continuous.  

In cases where the fluid ends at some extreme value of $x$, we always know the value of $Y$ at that boundary and use it as a boundary condition. One example is the case of a solid wall at the boundary; here, $Y$ vanishes on the wall since the Lagrangian displacement $\xxx$ vanishes. Another example is when a medium with constant total pressure exists outside the fluid, such as a hydrodynamic atmosphere with negligible density. In this case, $1/Y$ vanishes at the interface since the perturbation of the total pressure $\yyy$ vanishes. (We arrive at the same result if we consider a nonzero perturbation in the atmosphere, solve the problem on both sides while requiring the continuity of $Y$, and ultimately take the limit of zero density outside.)

The nontrivial boundary conditions that need closer examination correspond to cases where the fluid extends to theoretically infinite distances $x\to+\infty$ or $x\to-\infty$. At these distances the unperturbed fluid is homogeneous, and the solutions of the principal equation are given in Appendix~\ref{appendixB}.
These solutions correspond to two oppositely moving waves in the $\hat{x}$ direction, with one wave's amplitude increasing exponentially with $x$ and the other's decreasing.
To avoid the wave with the diverging amplitude,\footnote{There is a way to automatically find the non-diverging solution following the Schwarzian approach of Ref.~\cite{2024Symm...16..648V}. To obtain analytical expressions though, as we attempt here, we can directly find the asymptotic solutions.}
the solution of the principal equation should approach one of the constant values $Y_\pm=\dfrac{f_{11}\pm \sqrt{f_{11}^2+f_{12}f_{21}}}{f_{21}}
=\dfrac{f_{12}}{-f_{11}\pm \sqrt{f_{11}^2+f_{12}f_{21}}}$
that correspond to complex wavenumbers $\pm K=\pm i\sqrt{f_{11}^2+f_{12}f_{21}}$, respectively.  
Assuming that $\sqrt{f_{11}^2+f_{12}f_{21}}$ is the principal value of the root (with positive real part), we choose the upper sign at $x\to+\infty $ and the lower sign at $x\to-\infty $.
\\ In case that $\Im K=0$ (i.e., $f_{11}^2+f_{12}f_{21}$ is a negative real number), the wave's amplitude remains constant at infinity, and we can choose the solution whose sign of $\pm \Re K$ corresponds to the desired propagation. 

Note that $\sqrt{f_{11}^2+f_{12}f_{21}}=\sqrt{-\ktilde^2+\dfrac{\rho_0g'}{A}\ktilde^2+\dfrac{\rho_0^2g^2\bm k_0^2}{S}}
$.
In the presence of gravity this is not constant sine the unperturbed total pressure is variable with its gradient balancing gravity. However, in cases where the fluid extends to theoretically infinite distances (in which $S\to\infty$ since the total pressure also reaches theoretically infinite values), and uniform gravity, it approaches a constant $\sqrt{f_{11}^2+f_{12}f_{21}}\to|\bm k_0|$ corresponding to the incompressible limit.  

In the absence of gravity $\sqrt{f_{11}^2+f_{12}f_{21}}=\sqrt{-\ktilde^2}$ and without loss of generality we can choose the sign such that 
$\ktilde=\pm K=\pm i\sqrt{-\ktilde^2}$ (taking the principal value of the root and the upper/lower sign at $x\to\pm\infty$, respectively).

\subsubsection{Summary of equations and boundary conditions}  

The equations needed to apply the minimalist approach are summarized in Table~\ref{tab:formulas}.
\begin{table}[h!] \centering \begin{tabular}{ |m{1.72cm}  m{11.28cm}| } \hline 
		Principal equation: & \smallskip  \( Y' =f_{21} Y^2-2f_{11}Y-f_{12} \) 
		\smallskip \\ \hline  
		with & \smallskip  \( f_{11}=\dfrac{\rho_0 g}{A} \left(\dfrac{\rho_0\omega_0^2\F^2}{S}-\bm k_0^2\right) , 
		\quad f_{12}= \dfrac{\ktilde^2}{A} ,  \)
		\\ &
		\( f_{21}=-A-\dfrac{\rho_0^2g^2}{A}\left(\dfrac{\F^4}{S}-\bm k_0^2\right) +\rho_0g', \) 
		\\  
& \(\omega_0=\omega-\bm k_0\cdot\bm V_0 , \  \F =\bm k_0\cdot\bm B_0 , \
A=\rho_0\left(\omega_0^2-{\F^2}/{\rho_0}\right) ,
			 \) 
		\\
\multicolumn{2}{|c|}{  \(
	{S}/{\rho_0^2}= \left(c_s^2+{B_0^2}/{\rho_0}\right)\omega_0^2-c_s^2{\F^2}/{\rho_0}  , \
	\ktilde^2=\dfrac{\omega_0^4 }{S/\rho_0^2}-\bm k_0^2 , \
			\Pi_0'= -\rho_0g
			, \ \Pi_0=\dfrac{\rho_0 c_s^2}{\Gamma}+\dfrac{B_0^2}{2}
			. \)  }
		\\ \hline 
		\smallskip Boundary conditions: & \( Y  \) continuous everywhere, asymptotically \( Y|_{x=\pm \infty}=\dfrac{f_{11}\pm \sqrt{f_{11}^2+f_{12}f_{21}}}{f_{21}}
		. \)
		\\ \hline \end{tabular} \caption{Minimalist approach equations and boundary conditions} \label{tab:formulas} \end{table}
	
In addition, the following forms of the principal equation may be useful, especially if one looks for analytical solutions
\begin{eqnarray} 
	\dfrac{dY}{dx } =f_{21} \left(Y-\dfrac{f_{11}}{f_{21}}\right)^2-\dfrac{f_{11}^2+f_{12}f_{21}}{f_{21}}
	\Leftrightarrow  
	\dfrac{d }{dx } \left(\dfrac{1}{Y}\right) =f_{12}\left(\dfrac{1}{Y}+\dfrac{f_{11}}{f_{12}}\right)^2-\dfrac{f_{11}^2+f_{12}f_{21}}{f_{12}}
	\,,
\end{eqnarray} with 
\begin{eqnarray} 
 {f_{11}^2+f_{12}f_{21}}= {-\ktilde^2+ \dfrac{\ktilde^2}{A}\rho_0g' +\dfrac{\rho_0^2g^2\bm k_0^2}{S}} \,.
\end{eqnarray} 

Note that in cases with real oscillating eigenfunctions the $Y$ is real and becomes infinite at some points; this problem can be easily handled numerically by working with an angular variable $\arctan Y$ instead of $Y$ (or more generally with the $\arctan \dfrac{Y-{\cal Y}_2}{{\cal Y}_1}$ where ${\cal Y}_{1,2}$ are given functions of our choice), as described in Section 2 of Ref.~\cite{2024Symm...16..648V}. 

\section{MHD waves}\label{secmhdwaves}

The simpler case of disturbances inside a homogeneous magnetized plasma involves magnetohydrodynamic waves, which are analyzed in most plasma textbooks, e.g., Ref.~\cite{BS69}.

\subsection{Slow/fast magnetosonic waves}\label{secslowfast}
For homogeneous unperturbed state and zero gravity 
$f_{11}=0$, $f_{12}= \dfrac{\ktilde^2}{A} $, $  f_{21}=-A$, and the principal equation becomes $Y'=-AY^2-\dfrac{\ktilde^2}{A}$.

Its constant solutions are  $Y=i\dfrac{\ktilde}{A}$ (both signs are allowed).
According to equation~(\ref{eqsforxxxyyyY})
$\dfrac{\yyy'}{\yyy}=i\ktilde$, i.e., $ \ktilde$ is the $\hat x$ wavenumber. The solutions corresponds to the continuous spectrum of the slow/fast magnetosonic waves, and as already stated, the equation 
$\ktilde^2=\dfrac{\rho_0^2\omega_0^4}{S}-\bm k_0^2 $ is equivalent to the dispersion relation of these waves $ \dfrac{\omega_0^4}{\bm k_0^2+\ktilde^2}
-\omega_0^2\left(\dfrac{B_0^2}{\rho_0}+c_s^2\right)+ c_s^2\dfrac{\F^2}{ \rho_0}=0$.

The variable solution of the principal equation $Y=\dfrac{\ktilde}{A}\cot(\ktilde x+\phi_0)$ corresponds to a superposition of waves as discussed in Appendix~\ref{appendixB}.

\subsection{Alfv\`en waves}\label{secalfven}
The continuous spectrum of Alfv\`en waves $\omega_0^2=F^2/\rho_0$ corresponds to the singular case of the principal equation $A=0$. In that case $\xxx$ could be any function of $x$,
$\yyy= 0$ (and $Y\to\infty$), with finite   $\dfrac{\yyy}{A}=-\dfrac{\xxx'}{\ktilde^2} 
$ and $ \dfrac{1}{AY}=-\dfrac{\xxx'}{\ktilde^2\xxx}$.
(Note that for the Alfv\'en waves $\ktilde$ is given by $\ktilde^2=(\bm k_0\cdot \bm B_0/B_0)^2-\bm k_0^2=-(\bm k_0\times \bm B_0/B_0)^2$, and does not represent a wavenumber.)

\subsection{Two semi-infinite incompressible plasmas}
As another example on how to use the minimalist approach lets consider two semi-infinite incompressible, homogeneous, static, magnetized plasmas in the regimes $x>0$ (subscript "1") and $x<0$ (subscript "2"), and zero gravity.

With $S\to\infty$ in both regimes we have
$ f_{11}=0 $, $ f_{12}= -\dfrac{k_0^2}{A} $, $ f_{21}=-A  $.
The principal equation is
$Y' =\dfrac{k_0^2}{A}-AY^2$ and
has constant solutions $Y=\dfrac{f_{11}\pm \sqrt{f_{11}^2+f_{12}f_{21}}}{f_{21}}=\mp \dfrac{k_0 }{A}$.
According to the boundary conditions given in Table~\ref{tab:formulas} 
(or directly checking the sign of $\dfrac{\yyy'}{\yyy}=-f_{21}Y+f_{11} =\mp k_0$) the upper sign should be used for $x>0$ and the lower for $x<0$.  
Thus 
$Y=-\dfrac{k_0 }{\rho_1\omega^2-\F_1^2}$ for $x>0$ and $Y=+\dfrac{k_0 }{\rho_2\omega^2-\F_2^2}$ for $x<0$.

The continuity of $Y$ gives the dispersion relation 
$\omega^2=\dfrac{\F_1^2+\F_2^2}{\rho_1+\rho_2}$
representing a stable Alfv\'en wave whose amplitude drops exponentially as we move away from the contact discontinuity at $x=0$.
(We remind that $\F_1=\bm k_0\cdot \bm B_{01}$, $\F_2=\bm k_0\cdot \bm B_{02}$, and that all the unperturbed states considered in the paper have the form described in Section~\ref{seclinearanalysis}.)

\subsection{Effect of finite depth}
Following the previous example, suppose now that the bottom plasma has finite depth $H$ (there is a solid wall at $x=-H$).  
Then in the bottom part the solution of the principal equation $Y' =\dfrac{k_0^2}{A_2}-A_2Y^2$ that vanishes at $x=-H$ is 
$Y=\dfrac{k_0}{A_2}\tanh[k_0(x+H)]$. 
(In the upper part the accepted solution corresponding to vanishing amplitude at $x\to+\infty$ continues to be $Y=-\dfrac{k_0 }{A_1}$ as before.)

The continuity of $Y$ at $x=0$ gives the dispersion relation 
$\omega^2=\dfrac{\F_1^2+\F_2^2\coth(k_0H)}{\rho_1+\rho_2\coth(k_0H)}$.
Evidently the relation describes a stable Alfv\'en wave, modified by the presence of the wall at $x=-H$ and the reflections from that wall.

\section{Gravity waves and Rayleigh-Taylor instability}\label{secRayleigh-Taylor}
Density discontinuities in fluids within a gravitational field can drive gravity waves. These include disturbances at the interface between two fluids, such as between the atmosphere and the ocean. When a heavier fluid lies above a lighter fluid, the perturbation becomes unstable, leading to what is known as the Rayleigh-Taylor instability. This phenomenon was first described in the pioneering research papers by Rayleigh \cite{Rayleigh82} and Taylor \cite{Taylor50}.

\begin{figure}
	\includegraphics[align=t,width=0.45\textwidth]{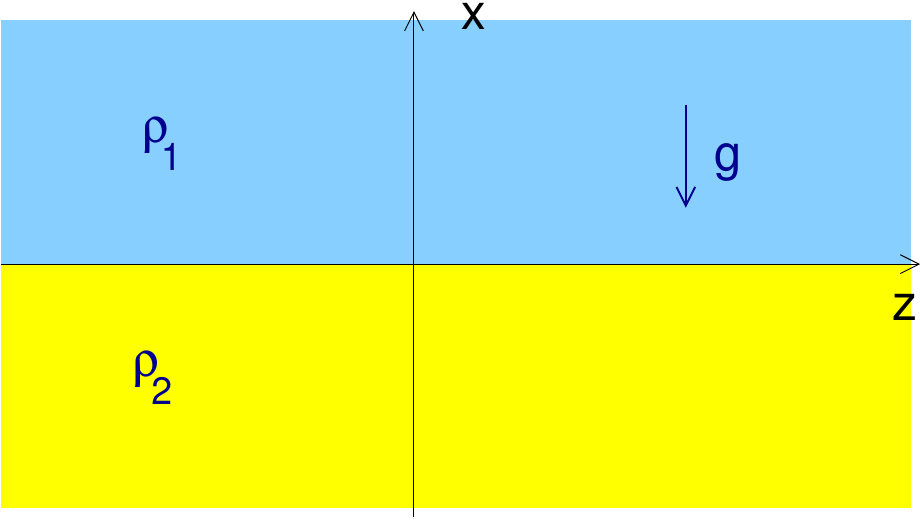}
	\hfill
	\includegraphics[align=t,width=0.45\textwidth]{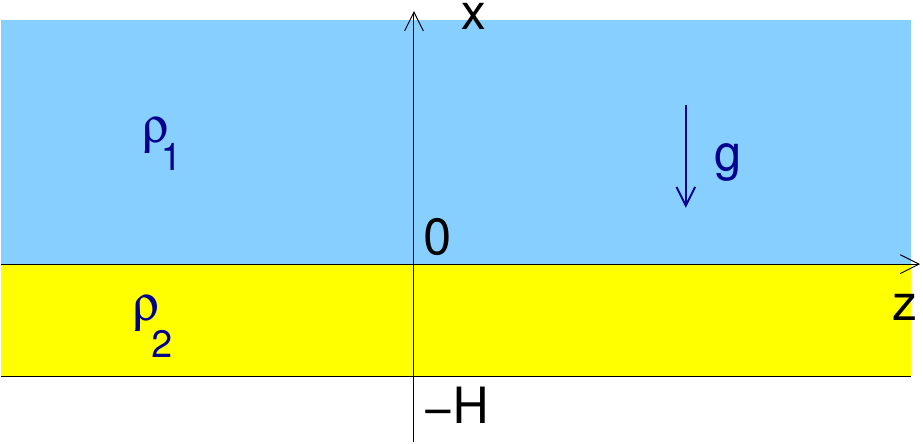} 
	\caption{The unperturbed state of two fluids in contact at $x=0$.
	Left panel: semi-infinite fluids. Right panel: the bottom part has finite depth $H$. 
		\label{figparker12}
	}
\end{figure}

Consider two semi-infinite incompressible, static, magnetized plasmas, the first (subscript "1") in the region $x>0$ and the second (subscript "2") in the region $x<0$, inside uniform gravity $-g\hat x$, as in the left panel of Figure~\ref{figparker12}. 

With $S\to\infty$ in both regimes we have
$ f_{11}=-g\rho_0\dfrac{k_0^2}{A} $, $ f_{12}= -\dfrac{k_0^2}{A} $, $ f_{21}=-A+g^2\rho_0^2\dfrac{k_0^2}{A} $.
The principal equation is
$Y' =\dfrac{k_0^2(1+g\rho_0 Y)^2}{A}-A Y^2$ and
has constant solutions 
$Y=\dfrac{f_{12}}{-f_{11}\pm \sqrt{f_{11}^2+f_{12}f_{21}}}= \dfrac{-k_0 }{\rho_0 gk_0\pm A}$.
(Although these parts are not homogeneous, since there must be a gradient of the total pressure balancing gravity, the incompressibility assumption allows to have constant solutions, provided that $\rho_0$ and $A$ are constant in both parts.) 

Choosing the correct sign in each part following Table~\ref{tab:formulas} (or directly using $\dfrac{\yyy'}{\yyy}=-f_{21}Y+f_{11} =\mp k_0$)  
we conclude that $Y=\dfrac{-k_0 }{\rho_1\omega^2-\F_1^2+\rho_1 gk_0}$ for $x>0$ and $Y=\dfrac{-k_0 }{-\rho_2\omega^2+\F_2^2+\rho_2 gk_0}$ for $x<0$.

The continuity of $Y$ gives the dispersion relation 
$\omega^2=-\dfrac{\rho_1-\rho_2}{\rho_1+\rho_2}gk_0+\dfrac{\F_1^2+\F_2^2}{\rho_1+\rho_2}$,
a result derived in Ref.~\cite{1961hhs..book.....C}.

This relation includes as subcases the surface  gravity waves corresponding to zero magnetic fields and $\rho_1\ll\rho_2$, for which $\omega^2=gk_0$, and the classical Rayleigh-Taylor instability corresponding to zero magnetic fields and heavier fluid on top of lighter, for which the growth rate is $\omega=i\sqrt{\dfrac{\rho_1-\rho_2}{\rho_1+\rho_2}gk_0}$.
It also shows the stabilizing effect of the magnetic field through its magnetic tension manifest in the dispersion relation in the presence of $F_1^2$ and $F_2^2$. It always reduces the growth rate, and if it is sufficiently strong it makes $\omega^2$ positive, converting the disturbance from an instability to an Alfv\'en wave. 

\subsection{Effect of finite depth}

If the bottom plasma has finite depth $H$ (there is a solid wall at $x=-H$, see the right panel of Figure~\ref{figparker12}) we need to consider the variable solution of the principal equation 
$Y=\dfrac{f_{12}}{-f_{11}-K\cot(Kx+\phi_0)} $ with $K=i\sqrt{f_{11}^2+f_{12}f_{21}}=ik_0$
(given in Appendix~\ref{appendixB}). Requiring $Y$ to vanish at $x=-H$ we find $Y=\dfrac{-k_0}{\rho_2gk_0-A_2\coth(k_0x+k_0H)}$ for $x<0$.

As a specific example suppose the upper part is a hydrodynamic medium with negligible density. In that case $Y=\infty$ for $x>0$ and the continuity of $Y$ gives the dispersion relation for gravity waves $\omega^2=gk_0\tanh(k_0H)+\dfrac{F_2^2}{\rho_2}$, modified by the presence of the magnetic field.
(The same can be found considering the constant solution $Y=\dfrac{-k_0 }{\rho_1\omega^2 +\rho_1 gk_0}$ in the upper part, requiring the continuity of $Y$ at $x=0$, and then taking the limit $\rho_1/\rho_2\to 0$.)
In the shallow water approximation $k_0H\ll 1$ we get $\omega^2=gHk_0^2+\dfrac{F^2}{\rho_2}$.

\section{Gravito-acoustic waves}\label{secgravitoacousticwaves}

In the examples presented up to now the fluids where homogeneous at least partially, allowing to easily solve the principal equation. In inhomogeneous cases we rely on the numerical integration of the principal equation. An exception to that, where it is possible to solve analytically the differential equation although the fluid is not homogeneous, is the case of gravito-acoustic waves, analyzed e.g., in Ref.~\cite{Goedbook2}. 
In addition it serves as a nice example showing the interplay between acoustic and gravity waves. 

For a hydrodynamic fluid between two solid walls at $x=0$ and $x=H$, inside homogeneous gravity $\bm g=-g\hat x$, with unperturbed density $\rho_0=\rho_{00}e^{-\alpha x}$, pressure $P_0=\dfrac{\rho_0g}{\alpha}$ satisfying the hydrostatic equilibrium condition $P_0'=-\rho_0 g$, and $\dfrac{1}{\alpha}=\dfrac{c_s^2}{\gamma g}$ the constant scaleheight,
we get $f_{11}=-\dfrac{gk_0^2}{\omega^2}$, 
$f_{12}=\dfrac{1/c_s^2-k_0^2/\omega^2}{\rho_{00}}e^{\alpha x}$,
$f_{21}=\rho_{00}(g^2k_0^2/\omega^2-\omega^2)e^{-\alpha x}$, and the principal equation can be written as
$$\dfrac{d(Y\rho_{00}\omega^2 e^{-\alpha x})}{dx}=\left(\dfrac{g^2k_0^2}{\omega^4}-1\right)(Y\rho_{00}\omega^2e^{-\alpha x})^2+\left(2\dfrac{gk_0^2}{\omega^2}-\alpha\right)(Y\rho_{00}\omega^2e^{-\alpha x})+k_0^2-\dfrac{\omega^2}{c_s^2} 
\,. $$
Its general solution is $\left(\dfrac{g^2k_0^2}{\omega^4}-1\right)Y\rho_{00}\omega^2e^{-\alpha x}=\dfrac{\alpha}{2}-\dfrac{gk_0^2}{\omega^2}+q\tan(qx+C)$
with $q=\sqrt{-\dfrac{\alpha^2}{4}+\dfrac{\omega^2}{c_s^2}-k_0^2+\dfrac{k_0^2N^2}{\omega^2} }$,
and $N^2=g\left(\dfrac{1}{\gamma P}\dfrac{dP}{dx}-\dfrac{1}{\rho}\dfrac{d\rho}{dx}\right)
=\alpha g-\dfrac{g^2}{c_s^2}
=\dfrac{\gamma-1}{\gamma}\alpha g$ the square of the Brunt-V\"ais\"al\"a (or buoyancy) frequency.
The boundary condition $Y|_{x=0}=0$ specifies $C$ and gives 
$\left(\dfrac{g^2k_0^2}{\omega^4}-1\right)Y\rho_{00}\omega^2e^{-\alpha x}=\dfrac{q^2+\left(\dfrac{\alpha}{2}-\dfrac{gk_0^2}{\omega^2}\right)^2}{q\cot(qx)+\dfrac{\alpha}{2}-\dfrac{gk_0^2}{\omega^2}}
=\dfrac{\left(\dfrac{g^2k_0^2}{\omega^4}-1\right)\left(k_0^2-\dfrac{\omega^2}{c_s^2}\right)}{q\cot(qx)+\dfrac{\alpha}{2}-\dfrac{gk_0^2}{\omega^2}}$.
The boundary condition $Y|_{x=H}=0$ requires 
$qH=n\pi$, $n=\pm 1, \pm 2, \dots$ and the dispersion relation is 
$-\dfrac{\alpha^2}{4}+\dfrac{\omega^2}{c_s^2}-k_0^2+\dfrac{k_0^2N^2}{\omega^2}=q^2
\Leftrightarrow \dfrac{\omega^4}{c_s^2}-\left(k_0^2+q^2+\dfrac{\alpha^2}{4}\right) \omega^2+ k_0^2N^2=0$. 
\\ The equation for $y_1$ is 
$\dfrac{y_1'}{y_1}=\dfrac{gk_0^2}{\omega^2}+\dfrac{k_0^2-\omega^2/c_s^2}{Y\rho_{00}\omega^2e^{-\alpha x}}=\dfrac{\alpha}{2}+q\cot(qx)$ with solution $y_1\propto e^{\alpha x/2}\sin(qx)$.
\\ (The cases $\omega^2=k_0^2c_s^2$ and $\omega^4=k_0^2g^2$ need to be examined separately. The former gives $q^2=-(\alpha/2-g/c_s^2)^2$ and the latter gives $\left(2\dfrac{gk_0^2}{\omega^2}-\alpha\right)(Y\rho_{00}\omega^2e^{-\alpha x})+k_0^2-\omega^2/c_s^2\propto \exp\left[\left(2\dfrac{gk_0^2}{\omega^2}-\alpha\right)x\right]$. Neither can satisfy the boundary conditions, so they cannot be accepted.)

\section{Kelvin-Helmholtz instability}\label{seckelvinhelmholtz}

Relative motion between two fluids induces an instability known as Kelvin-Helmholtz, as first described in the seminal papers by Lord Kelvin \cite{Thomson01111871} and Hermann von Helmholtz \cite{Helmholtz01111868}.

\subsection{Incompressible limit}
For the incompressible case and two semi-infinite plasmas, similarly to the Rayleigh-Taylor instability (Section~\ref{secRayleigh-Taylor}) we find that $Y=\dfrac{-k_0 }{\rho_1 gk_0+A_1}$ for $x>0$ and $Y=\dfrac{-k_0 }{ \rho_2 gk_0-A_2}$ for $x<0$.
The only difference is that $A_1=\rho_1(\omega-\bm k_0\cdot\bm V_{01})^2-\F_1^2$ and $A_2=\rho_2(\omega-\bm k_0\cdot\bm V_{02})^2-\F_2^2$ now depend on the velocities. 

The continuity of $Y$ gives the dispersion relation 
$\rho_1(\omega-\bm k_0\cdot\bm V_{01})^2+\rho_2(\omega-\bm k_0\cdot\bm V_{02})^2=\F_1^2+\F_2^2-(\rho_1-\rho_2)gk_0$
with solutions 
	\begin{eqnarray} 
\omega= \dfrac{\rho_1\bm V_{01}+\rho_2\bm V_{02}}{\rho_1+\rho_2}\cdot\bm k_0 
\nonumber \\
\pm i\sqrt{ 
	\dfrac{\rho_1\rho_2\left[(\bm V_{01}-\bm V_{02})\cdot\bm k_0\right]^2}{(\rho_1+\rho_2)^2}
	+\dfrac{\rho_1-\rho_2}{\rho_1+\rho_2}gk_0
	-\dfrac{(\bm k_0\cdot\bm B_{01})^2+(\bm k_0\cdot\bm B_{02})^2}{\rho_1+\rho_2}
} \,,
	\label{khdispincompr}
\end{eqnarray}
a result derived in Ref.~\cite{1961hhs..book.....C}.
The three terms within the square root have obvious physical meaning. The first is due to relative velocity and always leads to instability. The second is due to gravity and leads to stability/instability in the case of stable/unstable stratification (lighter fluid above heavier or the opposite).
The third term is due to the magnetic field stress and is always stabilizing.

\subsection{Compressible Kelvin-Helmholtz instability} 
Taking into account compressibility and neglecting gravity we have $f_{11}=0$, $ f_{12}= \dfrac{\ktilde^2}{A} $, $ f_{21}=-A$ and the constant solutions of the principal equation are
$Y=\dfrac{f_{11}\pm\sqrt{f_{11}^2+f_{12}f_{21}}}{f_{21}}=i\dfrac{\ktilde}{A}$ with $\ktilde=\pm i\sqrt{-\ktilde^2}$. According to Table~\ref{tab:formulas} we need to choose the upper sign in the $x>0$ and the lower in the $x<0$. 
Thus $\ktilde_1=i\sqrt{-\ktilde_1^2}$ and $Y=i\dfrac{\ktilde_1}{A_1}=\dfrac{\sqrt{-\ktilde_1^2}}{-A_1}$ for $x>0$, $\ktilde_2=-i\sqrt{-\ktilde_2^2}$ and $Y=i\dfrac{\ktilde_2}{A_2}=\dfrac{\sqrt{-\ktilde_2^2}}{A_2}$ for $x<0$. The dispersion relation is $\dfrac{ \ktilde_1}{A_1}=\dfrac{\ktilde_2}{A_2}$, or,
\begin{eqnarray} 
\dfrac{\sqrt{\bm k_0^2-\dfrac{\rho_1(\omega-\bm k_0\cdot \bm V_{01})^4 }{(\rho_1c_{s1}^2+B_{01}^2)(\omega-\bm k_0\cdot \bm V_{01})^2-(\bm k_0\cdot\bm B_{01})^2c_{s1}^2}}}{-\rho_1(\omega-\bm k_0\cdot \bm V_{01})^2+(\bm k_0\cdot\bm B_{01})^2}
=
\nonumber \\
\dfrac{\sqrt{\bm k_0^2-\dfrac{\rho_2(\omega-\bm k_0\cdot \bm V_{02})^4 }{(\rho_2c_{s2}^2+B_{02}^2)(\omega-\bm k_0\cdot \bm V_{02})^2-(\bm k_0\cdot\bm B_{02})^2c_{s2}^2}}}{\rho_2(\omega-\bm k_0\cdot \bm V_{02})^2-(\bm k_0\cdot\bm B_{02})^2} \,.
\label{khdispcompr}
\end{eqnarray}
(We remind that we need to take the principal values of both square roots, i.e., the ones with positive real parts.)

\subsection{On the physics of the Kelvin-Helmholtz instability} 

A relatively simple case for which we can find analytical results will help to understand the physics of the instability. 
We consider two hydrodynamic fluids with the same unperturbed characteristics 
(same densities $\rho_0$, same pressure $P_0$ as required from the equilibrium at the contact discontinuity, and same sound velocities $c_s$) and work in the frame where the two fluids move with opposite velocities $\pm V_0\hat z$ (with positive $V_0$ and the upper sign corresponding to $x>0$).

The dispersion relation becomes
$ \dfrac{\sqrt{1-M^2\left(\dfrac{\omega}{k_zV_0}-1\right)^2  }}{-\left(\dfrac{\omega}{k_zV_0}-1\right)^2 }
=\dfrac{\sqrt{1-M^2\left(\dfrac{\omega}{k_zV_0}+1\right)^2  }}{\left(\dfrac{\omega}{k_zV_0}+1\right)^2 } $,
where $M=\dfrac{V_0k_z/k_0}{c_s}$.
Unstable modes (with $\Im\omega>0$) exist for $M<\sqrt{2}$, with purely imaginary $\omega
=ik_zV_0\sqrt{\dfrac{2-M^2}{1+M^2+\sqrt{1+4M^2}}}$.
One way to find this result is to substitute $\omega=ik_zV_0\tan\dfrac{\mu}{2}$ when the dispersion relation reduces to $ \cos^2\mu+\cos\mu=M^2$ with solutions $\cos\mu=\dfrac{\sqrt{1+4M^2}-1}{2}$.
Actually the angle $\mu$ has an important meaning connected to the argument of
the resulting complex wavenumbers in the $\hat x$ direction, which are 
$\ktilde_{1,2}=-k_0e^{\mp i\mu}=-k_0\dfrac{\sqrt{1+4M^2}-1}{2}
\pm i k_0 \sqrt{\dfrac{\sqrt{1+4M^2}+1-2M^2}{2}}$.

Although the results apply for any angle between $\bm k_0$ and $\bm V_0$, to simplify the expressions from now on we restrict ourselves to the case $\bm k_0\parallel \bm V_0$, i.e., we consider a disturbance with $\bm k_0=k_0\hat z$ (with positive $k_0$).
The characteristics of the unstable mode, the Lagrangian displacement, 
and the perturbations of the pressure, density, and velocity (written in a way to show the phase difference between each quantity with $\xxx$) are 
\begin{eqnarray}
\omega=ik_0V_0\tan\dfrac{\mu}{2} \,, \quad \ktilde_{1,2}=-k_0e^{\mp i\mu} \,, \quad  \mu=\arccos\dfrac{\sqrt{1+4M^2}-1}{2}\,, \quad M=\dfrac{V_0}{c_s} \,, 
\\
\xxx e^{i(k_zz-\omega t)} \propto e^{k_0V_0 t \tan(\mu/2)} e^{\pm k_0x\sin\mu} e^{ik_0(z-x\cos\mu)}\,, \\
P_1 =\yyy= \dfrac{\rho_0k_0 V_0^2}{\cos^2(\mu/2)} e^{i\pi/2} \xxx \,, 
\quad  \rho_1 =\dfrac{P_1}{c_s^2} \,,
\\ 
V_{1x}= \dfrac{k_0V_0}{\cos(\mu/2)}e^{\pm i(\pi/2-\mu/2)} \xxx 
\,, \quad 
V_{1z}=\dfrac{k_0V_0}{\cos(\mu/2)} e^{\mp i(\pi/2-\mu/2)}\xxx
\,.
\end{eqnarray} 

Suppose we initially disturb the interface between the two fluids by $\Delta x|_{x=0,t=0}=D e^{ik_0z}$. The Lagrangian displacement along $\hat x$ at any later time and for any other fluid element has the form $\xi_x=\xxx e^{i(k_0z -\omega t)}=D e^{i(k_0z+\ktilde_{1,2} x-\omega t)}$; the goal is to find $\omega$ and $\ktilde_{1,2}$. Using the Lagrangian displacement $\bm\xi$ and the 
Lagrangian perturbations of the velocity, pressure, density, which are $\Delta\bm V=\dot{\bm \xi}$, $\Delta P$, $\Delta\rho=\dfrac{\Delta P}{c_s^2}$, respectively, 
the equations that will lead to that goal are the momentum $\rho_0\ddot{\bm \xi}=-\nabla (\Delta P) $ and the continuity $ \nabla\cdot\bm\xi=-\dfrac{\Delta\rho}{\rho_0}$
(coming from $\nabla\cdot\bm V=-\dfrac{\dot\rho}{\rho}$).
Since all perturbations are proportional to $e^{i(k_0z+\ktilde x-\omega t)}$ a time derivative has the meaning $\dfrac{\partial }{\partial t}+\bm V_0\cdot \nabla=-i(\omega\mp k_0V_0)$, and the divergence $\nabla=i\bm k_{t\ 1,2}$ with $\bm k_{t\ 1,2}=\ktilde_{1,2}\hat x+\bm k_0$ the total wavenumber in the two fluids
(their $\hat z$ components are the same, but the $\hat x$ components differ).

The momentum equation along $\hat x$ connects the displacement with the pressure perturbation $
\Delta P=\dfrac{\rho_0(\omega\mp k_0V_0)^2}{i\ktilde_{1,2}}\xxx$. Requiring $\Delta P$ and $\xxx$ to be continuous at the interface between the two fluids we arrive at a first relation between the unknowns $\dfrac{\ktilde_1}{A_1}=\dfrac{\ktilde_2}{A_2}$.

The pressure gradient along the interface is connected to the corresponding Lagrangian displacement $\xi_z$ through the momentum equation along $\hat z$, which gives $\xi_z=\dfrac{k_0}{\ktilde}\xxx$.
(The latter means that $\bm\xi\parallel \bm k_t$, a characteristic of longitudinal waves.)

The continuity also connects the two components of the displacement. It gives 
$k_0\xi_z+\ktilde \xxx=i\dfrac{\Delta P}{\rho_0 c_s^2}$.
Substituting $\Delta P=\dfrac{\rho_0(\omega\mp k_0V_0)^2}{i\ktilde_{1,2}}\xxx$ and $\xi_z=\dfrac{k_0}{\ktilde}\xxx$ we arrive at the second relation between the unknowns (one for each fluid) $ \ktilde_{1,2}^2=\dfrac{(\omega\mp k_0V_0)^2}{c_s^2}-k_0^2$.

The perturbation essentially consists of two sound waves in the two fluids with wavenumbers $\bm k_{t\ 1,2}=\ktilde_{1,2}\hat x+\bm k_0$. In the fluid rest frames the two waves move with different velocities satisfying
$(\omega-\bm k_0\cdot\bm V_0)^2=c_s^2\ktilde_{1,2}^2$, a relation equivalent to $ \ktilde^2=\dfrac{\rho_0^2\omega_0^4}{S}-k_0^2$ with $S=\rho_0^2c_s^2\omega_0^2$.
The two waves meet at the contact discontinuity and as a consequence they have the same $\bm k_0$ and $\omega$ such that their phases $\bm k_t\cdot (\bm r-\bm V_0t)-(\omega-\bm k_0\cdot\bm V_0) t=\bm k_t\cdot \bm r -\omega t$ are continuous at the interface.
As already presented, for the case of fluids with the same density, in a frame in which they move with opposite velocities $\pm V_0\hat z$, and $\bm k_0=k_0\hat z$ is parallel to the relative velocity, the resulting unstable mode has $\omega=ik_0V_0\tan\dfrac{\mu}{2}$ and
$\ktilde_{1,2}=-k_0e^{\mp i\mu}$, with $\mu=\arccos\dfrac{\sqrt{1+4M^2}-1}{2}$, $M=\dfrac{V_0}{c_s}$.

The value of the complex wavenumber $\ktilde$ (in each part) is directly connected to compressibility through $S$. 
In the incompressible limit $S\to\infty$ it is purely imaginary, since in that limit $M=0$, $\mu=\pi/2$ and $\ktilde=\pm i k_0$.
(Note that in the incompressible limit the continuity $\nabla\cdot\bm\xi=0$ gives $\bm k_t\cdot \bm \xi=0$. Since the two vectors are complex, this does not mean $\bm k_t\bot\bm\xi$. It rather means $\bm k_t^2=0$, using $\bm k_t\parallel \bm\xi$ from the momentum equation.
In addition, the relation $\bm k_t^2=0$ does not mean that the vector $\bm k_t$ is zero.)
If we consider cases with decreasing $S$, i.e., decreasing $c_s$ keeping $V_0$ the same, in which the compressibility becomes more and more important, the first part of the expression of $ \ktilde^2=\dfrac{\rho_0^2\omega_0^4}{S}-k_0^2$ affects the value of $\ktilde$ in two ways. Firstly the imaginary part decreases, meaning that the perturbation survives at longer distances from the interface. Secondly the $\Re\ktilde$ increases contributing more to the phase of the perturbation which is $k_0(z-x\cos\mu)$. 
Both effects are expressed through the angle $\mu$, which decreases with decreasing $S$ (increasing $M$). The growth rate is also connected to $S$ through $\mu$. It decreases as the result of compressibility, from $k_0V_0$ in the limit $S\to\infty$ to zero when $M=\sqrt{2}$, corresponding to the minimum $S$ for which the perturbation is unstable.

\begin{figure}
	\centering
	\includegraphics[width=.75\textwidth]{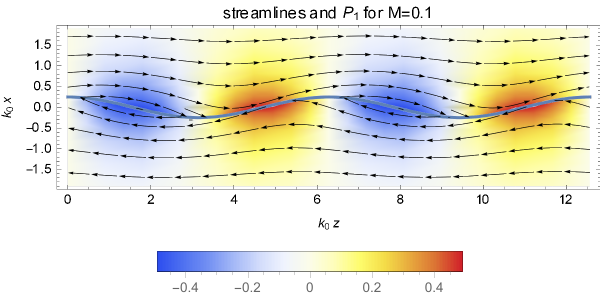}
	\\
	\includegraphics[width=.75\textwidth]{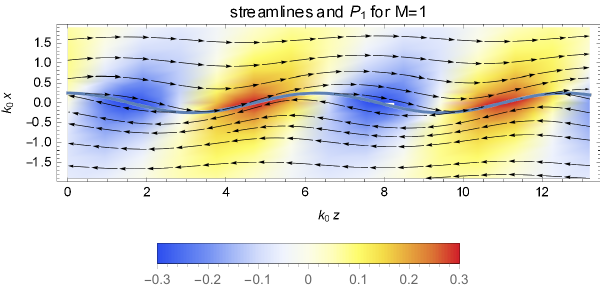}
	\\
	\includegraphics[width=.75\textwidth]{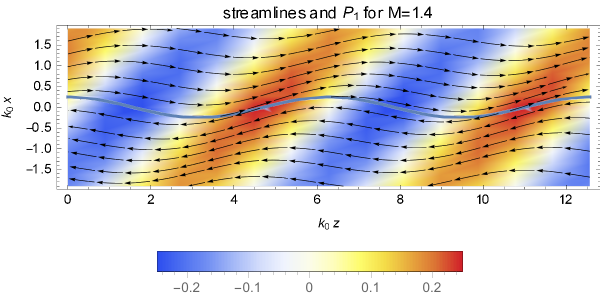} 
	\caption{The streamlines and the pressure perturbation for three cases, $M=0.1$ (upper panel), $M=1$ (middle panel), and $M=1.4$ (lower panel). 
		\label{fighydroKH}
	}
\end{figure} 

The vorticity is a key quantity in the Kelvin-Helmholtz instability. Its undisturbed value $\bm \zeta_0=-2V_0\delta(x) \hat y$ is the reason behind the development of the instability.
Even with the perturbation included it is nonzero only in the interface between the two fluids. Nevertheless, the motion of the fluids along the interface redistributes the vorticity compared to the unperturbed state.   
The related velocity inside the upper/lower fluid, just above/below the interface $\xi_x|_{x=0}=\xxx|_{x=0} e^{i(k_zz-\omega t)}$, is 
$V_{1z}|_{x=0^\pm}=k_0V_0 [\tan(\mu/2)\mp i]\xi_x|_{x=0}$.
The mean value $k_0V_0 \tan(\mu/2) \xi_x|_{x=0}$ shows that fluid accumulates near the positions of lower pressure, where the vorticity increases (in the $-\hat y$ direction) by $- ik_0\xi_x|_{x=0} \bm \zeta_0
$.
This accumulation of vorticity further increases the displacement $\xi_x|_{x=0}$ leading to instability. It becomes stronger for larger $\mu$, i.e., smaller $M$, since the mean value of $V_{1z}$ increases with $\mu$.

Three example solutions are shown in Figure~\ref{fighydroKH}.
The upper panel corresponds to an incompressible case with $M\ll 1$ for which $\omega\approx i (1-M^2)k_0V_0$ and $\ktilde_{1,2}\approx (-M^2\pm i)k_0$. 
(For $M\ll 1$ the approximate expression for $\mu$ is $\mu\approx \dfrac{\pi}{2}-M^2$.)
The exponential decrease of the perturbation as we move away from the interface is evident. 
The perturbed interface is shown, as also the circulation around points of minimum pressure
(the mean velocity of the fluids on the interface points toward these minimum pressure points). 
\\
The two other panels correspond to cases in which compressibility is important. The solution in the middle panel has $M=1$, $\omega= i 0.486k_0V_0$ and $\ktilde_{1,2}= (-0.618\pm i 0.786)k_0$
and in the lower panel $M=1.4$, $\omega= i 0.082k_0V_0$ and $\ktilde_{1,2}= (-0.987\pm i 0.163)k_0$.
The $\Re\bm k_t=\Re\ktilde\hat x+k_0\hat z$ has now a nonnegligible $\hat x$ part and as a result the iso-phase planes are $z-x\cos\mu=$ constant, tilted as shown in the panels.
\\
For cases approaching the maximum value of $M$ for which the perturbation is unstable
(the $M=\sqrt{2}$) the values of $\mu$, $\omega$ and $\Im\ktilde$ approach zero, and the perturbation practically consists of two standing sound waves (one in each fluid) with real wavenumbers.
(For $M^2-{2}\to 0^-$ the approximate expression for $\mu$ is $\mu\approx \sqrt{\dfrac{4-2M^2}{3}}$.) 

\subsection{Influence of magnetic field}
For simplicity we consider again two homogeneous fluids with the same unperturbed characteristics, and work in the frame where the fluids move with opposite velocities $\pm V_0\hat z$. 
Now there is also a constant magnetic field $\bm B_0=B_{0y}\hat y+B_{0z}\hat z$ in the unperturbed state, the same in both fluids.

Inspecting the dispersion relation~(\ref{khdispcompr}) we see that the magnetic field enters in two ways. Firstly through its pressure $B_0^2/2$, or equivalently the square of the Alfv\'en velocity $\bm v_{\rm A}=\dfrac{\bm B_0}{\sqrt{\rho_0}}$.
Secondly through its tension manifest in the terms $\bm k_0\cdot\bm B_0$, i.e. its component parallel to $\bm k_0$, or equivalently the component of the Alfv\'en velocity $v_{\rm A \parallel}=\dfrac{\bm B_0\cdot\bm k_0/k_0}{\sqrt{\rho_0}}$.

\begin{figure}
	\centerline{\includegraphics[width=.38\textwidth]{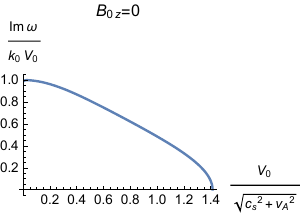} }
	\caption{The growth rate (normalized to $k_0V_0$) of the Kelvin-Helmholtz instability for two homogeneous fluids moving with $\pm V_0\hat z$, with same unperturbed density $\rho_0$, sound velocity $c_s$, magnetic field $B_{0y}\hat y$, and disturbance with $\bm k_0=k_0\hat z$. 
		\label{figallM}
	}

\vspace{10mm} 
	\includegraphics[width=.45\textwidth]{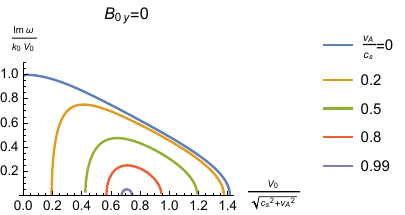} \hfill
	\includegraphics[width=.42\textwidth]{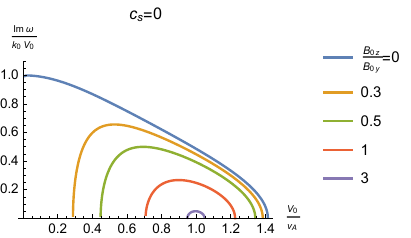} 
	\caption{Same as Figure~\ref{figallM}, but in the left panel for magnetic field $\bm B_0=B_{0z}\hat z$ along $\bm k_0=k_0\hat z$ with various strengths corresponding to the shown values of the ratio $v_{\rm A}/c_s$, and in the right panel for cold cases ($c_s=0$) and magnetic field  with various strengths and orientations corresponding to the shown values of the ratio $B_{0z}/B_{0y}$ (in all cases $\bm k_0=k_0\hat z$).
		\label{figallMz}
	}
\end{figure}

The magnetic pressure $B_0^2/2$ enters in the expression of $S$ and increases its value. Thus it moves the dynamics toward the incompressible limit, and according to the discussion of the previous section destabilizes
(the second terms inside the square roots equal $\dfrac{\rho_0^2\omega_0^4}{k_0^2S}$, so increase of $v_{\rm A}^2$ leads to the square roots being closer to unity). Actually if we include magnetic filed normal to $\bm k_0$ only, the dispersion relation is exactly equivalent to its hydrodynamic analogue, with the only difference that $M$ represents now the fast magnetosonic Mach number $M=\dfrac{V_0k_z/k_0}{\sqrt{c_s^2+v_{\rm A}^2}}$.
Figure~\ref{figallM} shows the resulting growth rates. Notably it includes as subcases the purely hydrodynamic case ($\bm v_A=0$) considered in the previous section, and the cold case ($c_s=0$) with only $B_{0y}$ component of the magnetic field.

The magnetic tension enters in the dispersion relation through $F=\bm k_0\cdot\bm B_0$ (essentially the component of the magnetic field along $\bm k_0$) in two places: inside $A$ and $S$.
It always has a stabilizing effect since the tension is a restoring force; we have seen that in the previous examples of Alfv\'en waves and the Rayleigh-Taylor instability, even in the incompressible limit. 
It also affects incompressibility through its appearance inside $S$. Since it decreases $S$ it moves the dynamics away from the incompressible limit, something that also in general stabilizes.
\\ Another related connection is that the magnetic tension affects the perturbation of the vorticity inside each fluid, which, using the expressions of the velocity perturbations given in Appendix~\ref{appendixA}, can be expressed as
$\Delta\bm\zeta=-i\dfrac{\omega_0 F\bm k_t\times\bm B_0}{\rho_0 A}\Delta\rho$.

In the following we present the methodology to find numerical results. 

The dispersion relation~(\ref{khdispcompr}) in the case under consideration can be written as
\begin{eqnarray} 
	\dfrac{\sqrt{1-\dfrac{\dfrac{V_0^2k_z^2/k_0^2}{c_s^2+v_{\rm A}^2}\left(\dfrac{\omega}{k_zV_0}-1\right)^4 }{\left(\dfrac{\omega}{k_zV_0}-1\right)^2-\dfrac{ c_s^2}{ c_s^2+v_{\rm A}^2}\dfrac{v_{\rm A \parallel}^2k_0^2}{V_0^2k_z^2}}}}{-\left(\dfrac{\omega}{k_zV_0}-1\right)^2+\dfrac{v_{\rm A \parallel}^2k_0^2}{V_0^2k_z^2}}
	=\dfrac{\sqrt{1-\dfrac{\dfrac{V_0^2k_z^2/k_0^2}{c_s^2+v_{\rm A}^2}\left(\dfrac{\omega}{k_zV_0}+1\right)^4 }{\left(\dfrac{\omega}{k_zV_0}+1\right)^2-\dfrac{ c_s^2}{ c_s^2+v_{\rm A}^2}\dfrac{v_{\rm A \parallel}^2k_0^2}{V_0^2k_z^2}}}}{\left(\dfrac{\omega}{k_zV_0}+1\right)^2-\dfrac{v_{\rm A \parallel}^2k_0^2}{V_0^2k_z^2}} \,,
	\label{khdispcomprB}
\end{eqnarray}
and can be solved numerically.
However, for the case of fluids with the same characteristics considered here it is possible to proceed analytically. 
As shown in Appendix~\ref{appendixC} the dispersion relation can be transformed to the following quartic polynomial equation for the square of the growth rate $\dfrac{(\Im\omega)^2}{k_0^2(c_s^2+v_{\rm A}^2)}=M^2\tan^2\dfrac{\Lambda}{2}$ 
\begin{eqnarray} 
		M^8 \tan^8\dfrac{\Lambda}{2}
		+2(1+2M^2)M^6\tan^6\dfrac{\Lambda}{2}
		\nonumber \\ 
		+[2M^2(1+3M^2)+2(2\cos^2\etamine+1)\cos^2\Theta-(2\cos^2\etamine+1)\cos^4\Theta]M^4\tan^4\dfrac{\Lambda}{2}
		\nonumber \\ 
		+\left\{ 2M^4(2M^2-1)-4M^2(2\cos^2\etamine-1)\cos^2\Theta
		\right. \nonumber \\ \left.
		+2\left[\cos^4\etamine-2(M^2-1)\cos^2\etamine-M^2\right]\cos^4\Theta
		-2\cos^2\etamine\cos^6\Theta\right\}M^2\tan^2\dfrac{\Lambda}{2}
		\nonumber \\
		+M^6(M^2-2)+2M^4(2\cos^2\etamine+1)\cos^2\Theta
		\nonumber \\
		-M^2[2\cos^4\etamine+2(M^2+2)\cos^2\etamine+M^2]\cos^4\Theta
		+2(M^2+\cos^2\etamine)\cos^2\etamine\cos^6\Theta
		=0
		\,, \label{khdispcomprBlambda}
\end{eqnarray}  
using the parametrization \begin{eqnarray} 
\omega=ik_zV_0\tan\dfrac{\Lambda}{2}
\,, \quad 
M=\dfrac{V_0k_z/k_0}{\sqrt{c_s^2+v_{\rm A}^2}}
\,, \quad 
\dfrac{v_{\rm A}}{c_s}=\tan\etamine
\,, \quad 
\dfrac{v_{\rm A \parallel}}{\sqrt{c_s^2+v_{\rm A}^2}}=\cos\Theta
 \,.
\end{eqnarray}
(Note that the angle $\Theta$ is connected to the angle between the magnetic field and the wavenumber since $\cos\Theta=\dfrac{\bm B_0\cdot \bm k_0}{B_0k_0}\sin\etamine$.)

The left panel of Figure~\ref{figallMz} shows the growth rate for cases where the magnetic field has only component along $\bm k_0$. In general the field decreases the growth rate and if it is sufficiently strong completely suppresses the instability.
Similar behavior is shown in the right panel of Figure~\ref{figallMz} for the cold case. For all strengths of the magnetic field, if its orientation is sufficiently close to the wavevector the magnetic tension completely suppresses the instability.  

\subsection{Range of instability}
It is interesting to explore the regions of $M$ for which the Kelvin-Helmholtz instability is present, shown in Figure~\ref{figallMz}.
The question is under what conditions the dispersion relation~(\ref{khdispcomprB}) has purely imaginary roots. For simplicity, we consider a disturbance with $\bm k_0=k_0\hat z$ parallel to the velocities. The results can be easily generalized. 

One might naively think that the extreme values of these instability regions can be found by setting $\omega=0$ in the dispersion relation (\ref{khdispcomprB}) and requiring both numerators to vanish (since the denominators are opposite real numbers). However, this implies $\tilde{k}=0$, resulting in no $x$-dependence of the disturbance. This case corresponds to magnetosonic waves in the frame of each fluid with total wavevectors $\bm k_t=k_0\hat z$ and $\omega_0=\mp k_0V_0$ with $V_0^4-V_0^2(v_{\rm A}^2+c_s^2)+c_s^2v_{\rm A \parallel}^2=0$ (see Section~\ref{secslowfast}). The absence of $x$ dependence makes these cases unrelated to the unstable modes. 

However, there are two other possibilities.
The vanishing of both denominators in the dispersion relation~(\ref{khdispcomprB}), for $\omega=0$ needs to be considered as a possible limiting case. This corresponds to Alfv\'en waves with $\omega_0=\mp k_0V_0$ with $ V_0^2=v_{\rm A \parallel}^2$ (see Section~\ref{secalfven}). Note that $\ktilde$ does not appear in the dispersion relation. Nevertheless, its value is nonzero, given by $\ktilde^2=k_0^2\left(1-\dfrac{v_{\rm A \parallel}^2}{v_{\rm A}^2}\right)$. The $x$-dependence of the disturbance allows for a possible connection with unstable modes. 

A third possibility is that the limiting values may correspond to bifurcations of the dispersion relation. It is actually evident from inspecting Figure~\ref{figallM} that the slope $\dfrac{\partial\omega}{\partial M}$ becomes infinity when $\omega=0$.
In general, a dispersion relation depends on various parameters, and the slope is defined as the derivative with respect to one of these parameters, while keeping all others constant.
For a dispersion relation of the form $f(\omega,p_1,p_2,\dots)=$ constant its differential 
$\dfrac{\partial f}{\partial \omega}d\omega +\displaystyle\sum_{n}\dfrac{\partial f}{\partial p_n}dp_n=0$
shows that bifurcation corresponds to $\dfrac{\partial f}{\partial \omega}=0$. 
Thus, extreme values of the instability regions may be connected to the condition $\left.\dfrac{\partial f}{\partial \omega}\right|_{\omega=0}=0$.
\\
In our case there are three parameters, and we can choose $p_1=M^2=\dfrac{V_0^2}{c_s^2+v_{\rm A}^2}$, $p_2=\dfrac{c_s^2}{c_s^2+v_{\rm A}^2}$, $p_3=\dfrac{v_{\rm A \parallel}^2}{V_0^2}$.
The algebra  gives that $\left.\dfrac{\partial f}{\partial \omega}\right|_{\omega=0}=0$ leads to 
$p_1-2+p_1p_3-2p_1p_2p_3^2-2p_2^2p_3^2+4p_2p_3=0$, or the following cubic for $V_0^2$
\begin{eqnarray}  
V_0^6 
-(2c_s^2+2v_{\rm A}^2-v_{\rm A \parallel}^2)V_0^4  
+\dfrac{2c_s^2v_{\rm A \parallel}^2}{c_s^2+v_{\rm A}^2}(2c_s^2+2v_{\rm A}^2-v_{\rm A \parallel}^2)V_0^2
-\dfrac{2c_s^4v_{\rm A \parallel}^4}{c_s^2+v_{\rm A}^2} 
=0 \,,
	\label{eqbifurcation}
\end{eqnarray} 
or,
$ M^6 
+(\cos^2\Theta-2) (M^4  
-2 \cos^2\etamine \cos^2\Theta M^2)
= 2 \cos^4\etamine \cos^4\Theta $.

For the left panel of Figure~\ref{figallMz} we have $\cos\Theta=\sin\etamine$, $\dfrac{v_{\rm A}}{c_s}=\tan\etamine$, and the extreme values of $M$ are given by $ M^6 
-(2 - \sin^2\etamine) M^4  
+ 2 \cos^2\etamine \sin^2\etamine(2 - \sin^2\etamine ) M^2
- 2 \cos^4\etamine \sin^4\etamine
=0
$.
The roots of this cubic are $M=\sin\etamine$, $M=\cos\etamine\sqrt{1-\sqrt{\cos(2\etamine)}}$, $M=\cos\etamine\sqrt{1+\sqrt{\cos(2\etamine)}}$, and the corresponding wavenumbers $\ktilde^2=0$, $k_0^2\cos(2\etamine)$, $k_0^2\cos(2\etamine)$, respectively. Thus the value $M=\sin\etamine$ corresponds to a magnetosonic wave without $x$ dependence (unrelated to instability), while the other two solutions $M=\cos\etamine\sqrt{1-\sqrt{\cos(2\etamine)}}$, $M=\cos\etamine\sqrt{1+\sqrt{\cos(2\etamine)}}$ are the extreme values of $M$ related to the unstable modes.
\\
The result is that for $\bm B_{0}\parallel \bm k_0$
the Kelvin-Helmholtz instability occurs only if $v_{\rm A}<c_s$ and for velocities in the interval  
$c_s\sqrt{1-\sqrt{\dfrac{c_s^2-v_{\rm A}^2}{c_s^2+v_{\rm A}^2}}}<V_0<c_s\sqrt{1+\sqrt{\dfrac{c_s^2-v_{\rm A}^2}{c_s^2+v_{\rm A}^2}}}$.
As we approach the limits the wavenumbers $\ktilde$ approach real values and the instability is transformed to magnetosonic waves with constant amplitudes and wavenumbers $\bm k_t=k_0\hat z \pm k_0 \sqrt{\dfrac{c_s^2-v_{\rm A}^2}{c_s^2+v_{\rm A}^2}} \hat x$.

For the right panel of Figure~\ref{figallMz} we have $\etamine=\pi/2$ and $\dfrac{B_{0z}}{B_{0y}}=\cot\Theta$ and the maximum value of $M$ corresponds to bifurcation for which 
$ M^6 -(1+\sin^2\Theta) M^4 =0$.
The nontrivial root is
$M=\sqrt{2-\dfrac{B_{0z}^2}{B_0^2}}$, and the corresponding wavenumbers 
$\ktilde^2=k_0^2\left(1-\dfrac{B_{0z}^2}{B_0^2}\right)$.
\\ The lower limit of the instability region corresponds to 
Alfv\'en waves $ V_0^2=v_{\rm A \parallel}^2$ so $M=\dfrac{|B_{0z}|}{B_0}$.
\\
The result is that for the cold case $c_s=0$
the Kelvin-Helmholtz instability occurs only if 
$\dfrac{|B_{0z}|}{\sqrt{\rho_0}}<V_0<\sqrt{ \dfrac{2B_0^2-B_{0z}^2}{\rho_0}}$.
As we approach the limits the wavenumbers $\ktilde$ approach real values and the instability is transformed to waves (Alfv\'en wave in the lower limit and magnetosonic in the upper) with constant amplitudes and wavenumbers $\bm k_t=k_0\hat z \pm k_0 \sqrt{1-\dfrac{B_{0z}^2}{B_0^2}} \hat x$.

\subsection{Bifurcations in the general case}
The dispersion relation in the general case is $f=\dfrac{A_2\ktilde_1}{A_1\ktilde_2}=1$, and taking its logarithmic derivative we conclude that 
bifurcations occur when $\dfrac{\partial\ktilde^2}{\ktilde^2\partial\omega}-2\dfrac{\partial A}{A\partial\omega}$ is continuous at the interface.
Equivalently, since $\dfrac{\ktilde}{A}$ is also continuous, $\dfrac{\partial\ktilde^2}{A^2\partial\omega}-2\ktilde^2\dfrac{\partial A}{A^3\partial\omega}$ is continuous.
Substituting $S$ and $A=\rho_0\omega_0^2-F^2$ this quantity equals 
\\
$
\dfrac{4\rho_0\omega_0}{A^3}\bm k_0^2
-\dfrac{4\rho_0\omega_0^3F^2}{A^3(Ac_s^2+\omega_0^2B_0^2)} 
-\dfrac{2\rho_0\omega_0^5(\rho_0c_s^2+B_0^2)}{A^2(Ac_s^2+\omega_0^2B_0^2)^2} 
$.

In the hydrodynamic case the latter simplifies to
$ 2\dfrac{2\bm k_0^2c_s^2-\omega_0^2}{\rho_0^2c_s^2\omega_0^5} $,
while in the cold case to $2\omega_0\dfrac{2\bm k_0^2v_{\rm A}^2-k_0^2v_{\rm A \parallel}^2-\omega_0^2}{\rho_0^2v_{\rm A}^2(\omega_0^2-k_0^2v_{\rm A \parallel}^2)^3} $.

\section{Summary}
The primary objective of this paper is to illustrate how the minimalist approach, as introduced in Ref.~\cite{2024Univ...10..183V}, can be applied to stability problems in planar geometry using Cartesian coordinates. This method highlights the approach's efficacy in determining the dispersion relation by integrating a single first-order differential equation, the \emph{principal equation}.

Although the mathematical formalism is quite cumbersome, it can be simplified by defining intermediate quantities with important physical meanings. For example, the complex $\ktilde$ has a direct connection to wave propagation and simultaneous amplitude variation. The function $S$ is related to compressibility, which depends on both thermal and magnetic pressure, and shows how the wave propagation and the growth rate of the instability depend on these factors. 
The function $F$ signifies the stabilizing nature of magnetic tension. All these aspects are discussed and reviewed when applying the method to classical instabilities.

A more extensive analysis was conducted on the Kelvin-Helmholtz instability, resulting in new findings. Specifically, analytical solutions of the dispersion relation in certain cases were derived, along with a study of bifurcations and their connection to the ranges of instability. These tools facilitate parametric studies, which are far from being considered complete even in classical instabilities. 

Detailed studies of specific problems in hydrodynamics or magnetohydrodynamics will benefit from the examples and formalism presented, as well as from the ideas on how to explore the possible existence of analytical solutions and specify the range of instability. Applications in other geometries and more general theoretical frameworks are also possible and will be presented in future works.
 
\vspace{6pt} 




\funding{This research received no external funding.}

\dataavailability{This research is analytical; no new data were generated or analyzed. If needed, more details on the study and the numerical results will be shared on reasonable request to the author.} 


\conflictsofinterest{The author declares no conflicts of interest.} 


%


\appendixtitles{yes} 
\appendixstart
\appendix


\section{Linearization}\label{appendixA}

The linearized equations (\ref{mass}--\ref{induction}) are (for $\omega_0\neq 0$)
\begin{eqnarray}
	\omega_0 \dfrac{\Delta\rho}{\rho_0} =-i V_{1x}' +k_z V_{1z}+k_yV_{1y} \,,
	\label{mass1}
	\\
	P_1-c_s^2\Delta\rho
	+i\dfrac{V_{1x}}{\omega_0} P_0'  
	=0 \,,
	\label{energy1}
	\\ -i\omega_0\rho_0V_{1x}=
	-\Pi_1'+i\F B_{1x}
	-\rho_1g
	\label{momentumvarpi1}
	\\
	-i\omega_0\rho_0V_{1y}+\rho_0V_{1x} V_{0y}'=
	-ik_y\Pi_1+i\F B_{1y}+B_{1x} B_{0y}'
	\label{momentumphi1}
	\\
	-i\omega_0\rho_0V_{1z}+\rho_0 V_{1x} V_{0z}'=
	-ik_z\Pi_1+i\F B_{1z}+B_{1x} B_{0z}'
	\label{momentumz1}
	\\ 
	\omega_0 B_{1x}= -\F  V_{1x}
	\,, \label{inductionvarpi1}
	\\
	\omega_0 B_{1y}=\omega_0B_{0y}\dfrac{\Delta\rho}{\rho_0}
	-\F V_{1y}  
	-i V_{1x} B_{0y}'
	+iB_{1x} V_{0y}'
	\,, \label{inductionphi1}
	\\ 
	\omega_0 B_{1z}=\omega_0B_{0z}\dfrac{\Delta\rho}{\rho_0}-\F V_{1z} 
	-i V_{1x} B_{0z}' +iB_{1x} V_{0z}'
	\,, \label{inductionz1}
\end{eqnarray}  
with 
$\F =\bm k_0\cdot \bm B_0$, 
$c_s$ the sound velocity (only the unperturbed is needed and is assumed known function of $x$),
$P_0=\Pi_0-\dfrac{B_0^2}{2}$,
$P_1=\Pi_1-B_{0y}B_{1y}-B_{0z}B_{1z}$,
and $\Delta\rho=\rho_1+\dfrac{iV_{1x}}{\omega_0} \rho_0'$.
In the above expressions the relation $\nabla\cdot \bm B=0$ was used. This can be seen as a consequence of the induction equation~(\ref{induction}), but shows more directly that the relation $ B_{1x}'+ik_yB_{1y}+ik_zB_{1z}=0$ holds.

We can use the equations (\ref{momentumphi1})--(\ref{inductionz1}) of the system,
together with the definitions of $\Delta\rho=\rho_1+\dfrac{iV_{1x}}{\omega_0}\rho_0'$ and $\xxx=\dfrac{i  V_{1x}}{\omega_0}$, to express the perturbations of the density, velocity and the magnetic field as functions of $\Delta\rho$, $\xxx$, and $\Pi_1$: 
\begin{eqnarray} 
	\rho_1=-\rho_0' \xxx +\Delta\rho
	\,, \\
	V_{1x} =-i\omega_0\xxx  \,, 
	\\
	AV_{1y}= -A V_{0y}' \xxx+
	k_y\omega_0\Pi_1-\F \omega_0B_{0y}\dfrac{\Delta\rho}{\rho_0}
	\,, \label{momentumphi2}
	\\
	AV_{1z}=-A V_{0z}' \xxx+
	k_z\omega_0\Pi_1-\F \omega_0B_{0z}\dfrac{\Delta\rho}{\rho_0}
	\,, \label{momentumz2} \\
	B_{1x}= i\F \xxx
	\,, \label{inductionvarpi2}
	\\
	AB_{1y}= -A B_{0y}' \xxx-\F 
	k_y \Pi_1 +\omega_0^2B_{0y} \Delta\rho 
	\,, \label{inductionphi3}
	\\ 
	AB_{1z}=-A B_{0z}' \xxx  -\F  k_z \Pi_1 +\omega_0^2B_{0z} \Delta\rho 
	\,, \label{inductionz3}
\end{eqnarray} 
where $A=\rho_0\omega_0^2-\F^2$.

The substitution of $V_{1y}$, $V_{1z}$ in equation (\ref{mass1}) gives
\begin{eqnarray}
	\omega_0^2 \Delta\rho= -A \xxx'+
	\left(k_y^2+k_z^2\right)\Pi_1 
	\,, \label{mass4} 
\end{eqnarray}
while the remaining equations (\ref{energy1}), (\ref{momentumvarpi1}) are
\begin{eqnarray} 
	\Pi_1-c_s^2\Delta\rho+\xxx P_0' = B_{0y} B_{1y}+ B_{0z}B_{1z}\,, 
	\label{energy4}
	\\ 
	0=
	-\Pi_1'-\rho_1g+A \xxx 
	\,. \label{momentumvarpi4} 
\end{eqnarray} 

We introduce the perturbation of the total pressure in the perturbed position $\yyy=\Pi_1+\xxx \Pi_0' =\Pi_1-\rho_0 g\xxx$, 
a quantity that is continuous everywhere (similarly to $\xxx$).

The substitution of $B_{1y}$, $B_{1z}$ in equation (\ref{energy4}), using also the equilibrium of the unperturbed state
$ P_0' +B_{0y} B_{0y}' +B_{0z} B_{0z}'=-\rho_0 g$, gives
\begin{eqnarray}
	\Delta\rho=\rho_0^2
	\dfrac{\omega_0^2\yyy+g\F^2\xxx}{S}
	\,,
\end{eqnarray} 
where $S=\rho_0(Ac_s^2+\omega_0^2B_0^2)$.

Substituting this $\Delta\rho$ and $\rho_1=\Delta\rho-\xxx\rho_0'$
in equations (\ref{mass4}) and (\ref{momentumvarpi4}) we arrive at the system~(\ref{eqsystemxxxyyy}). 

After solving this system and find $\xxx$, $\yyy$ we return to the rest of the equations and find all other perturbations. We summarize the expressions below
(in which the left-hand sides represent the Lagrangian perturbation of each quantity, i.e., the perturbation in the perturbed position).
\begin{eqnarray}
	V_{1x}=-i\omega_0\xxx \,, \label{vx5}
	\\ 
	V_{1y}+ V_{0y}' \xxx=
	\omega_0k_y\dfrac{\yyy+\rho_0 g \xxx}{A}- \rho_0\omega_0\F B_{0y}\dfrac{\omega_0^2\yyy+g\F^2\xxx}{AS} 
	\,, \label{momentumphi5}
	\\
	V_{1z}+ V_{0z}' \xxx=
	\omega_0k_z\dfrac{\yyy+\rho_0 g \xxx}{A}- \rho_0\omega_0\F B_{0z}\dfrac{\omega_0^2\yyy+g\F^2\xxx}{AS} 
	\,, \label{momentumz4}
	\\
	B_{1x}= i\F \xxx
	\,, \label{inductionvarpi5}
	\\
	B_{1y}+ B_{0y}' \xxx= -\F 
	k_y \dfrac{\yyy+\rho_0 g \xxx}{A}+\rho_0^2\omega_0^2B_{0y}\dfrac{\omega_0^2\yyy+g\F^2\xxx}{AS} 
	\,, \label{inductionphi5}
	\\ 
	B_{1z}+ B_{0z}' \xxx=-\F  k_z \dfrac{\yyy+\rho_0 g \xxx}{A}+\rho_0^2\omega_0^2B_{0z}\dfrac{\omega_0^2\yyy+g\F^2\xxx}{AS} 
	\,, \label{inductionz5} \\
	\rho_1+\rho_0'\xxx=
	\rho_0^2\dfrac{\omega_0^2\yyy+g\F^2\xxx}{S} \,, 
	\label{energy5}
	\\ 
	P_1+ P_0' \xxx=c_s^2\rho_0^2\dfrac{\omega_0^2\yyy+g\F^2\xxx}{S} 
	\,. \label{pressure5}
\end{eqnarray}

\section{Solutions of the principal equation in the homogeneous case}\label{appendixB}

If the array elements $f_{ij}$ are constants the principal equation has two kind of solutions. 
Either constants $Y_\pm$ satisfying $0=f_{21} Y_\pm^2-2f_{11}Y_\pm-f_{12} \Leftrightarrow Y_\pm=\dfrac{f_{11}\mp iK}{f_{21}} 
$ with $K=i\sqrt{f_{11}^2+f_{12}f_{21}}$, or variable $Y=\dfrac{f_{11}-K\cot(Kx+\phi_0)}{f_{21}} $, with $\phi_0$ a constant of integration.

Each one of the former corresponds to one way wave propagation in the $\hat x$ (or $-\hat x$) direction and the latter to a superposition of these two waves. 
This becomes clear if we find $\yyy$ and $\xxx$ using equations~(\ref{eqsforxxxyyyY}). It is even simpler to look for solutions of the system~(\ref{eqsystemxxxyyy}), which is linear with constant coefficients in this case and admits solutions of the form $y_{1,2}\propto e^{\pm iKx}$ with $\dfrac{\xxx}{\yyy}=Y_\pm$.
The general solution is 
$\yyy=C_+e^{iKx}+C_-e^{-iKx}$, $\xxx=C_+Y_+e^{iKx}+C_-Y_-e^{-iKx}=\dfrac{f_{11}}{f_{21}}\yyy-i\dfrac{K}{f_{21}}C_+e^{iKx}+i\dfrac{K}{f_{21}}C_-e^{-iKx}$. 

We can simplify the expressions of the eigenfunctions to 
$\yyy=2iD\sin(Kx+\phi_0)$, $\xxx=2iD\dfrac{f_{11}\sin(Kx+\phi_0)-K\cos(Kx+\phi_0)}{f_{21}}$, substituting $C_\pm=\pm De^{\pm i\phi_0}$, and their ratio agrees with the solution of the principal equation given above  $Y=\dfrac{\xxx}{\yyy}= \dfrac{f_{11}-K\cot(Kx+\phi_0)}{f_{21}} $.
Note that the last expression for $Y$ can also be written as $\dfrac{1}{Y}=\dfrac{\yyy}{\xxx}= \dfrac{-f_{11}-K\cot(Kx+\phi_0')}{f_{12}}$ with $C_\pm Y_\pm=\pm D'e^{\pm i\phi_0'}
\Leftrightarrow \dfrac{Y_+}{Y_-}=e^{2i(\phi_0'-\phi_0)}$, and we get the equivalent expressions of the eigenfunctions $\xxx=2iD'\sin(Kx+\phi_0')$, $\yyy=2iD'\dfrac{-f_{11}\sin(Kx+\phi_0')-K\cos(Kx+\phi_0')}{f_{12}} $.

The constant $Y$ solutions correspond to $C_\mp=0\Leftrightarrow \phi_0=\mp i\infty$, giving $\yyy\propto e^{\pm i Kx}$, $Y= \dfrac{f_{11}\mp iK}{f_{21}}=\dfrac{-f_{12}}{f_{11}\pm iK}$.

The complex wavenumbers of the two waves are $\pm K=\pm i\sqrt{f_{11}^2+f_{12}f_{21}}$. 
Taking the principal value of the square root (with positive real part), the upper/lower sign corresponds to a wave whose amplitude decreases with increasing/decreasing $x$.

\section{Analytical solutions of the magnetized Kelvin-Helmholtz instability}\label{appendixC} 

With the parametrization $\omega=ik_zV_0\tan\dfrac{\Lambda}{2}$, $\dfrac{V_0k_z/k_0}{\sqrt{c_s^2+v_{\rm A}^2}}=M$, 
$\dfrac{v_{\rm A}}{\sqrt{c_s^2+v_{\rm A}^2}}=\sin\etamine$, $\dfrac{c_s}{\sqrt{c_s^2+v_{\rm A}^2}}=\cos\etamine$,
$\dfrac{ v_{\rm A \parallel}}{ \sqrt{c_s^2+v_{\rm A}^2}}=\cos\Theta $, the dispersion relation~(\ref{khdispcomprB})
becomes   
\begin{eqnarray} 
		\dfrac{i\sqrt{1-\dfrac{M^4e^{-2i\Lambda}/\cos^2\dfrac{\Lambda}{2} }{M^2e^{-i\Lambda}-\cos^2\dfrac{\Lambda}{2}\cos^2\etamine\cos^2\Theta}}}{\dfrac{M^2 e^{-i\Lambda}}{\cos^2\dfrac{\Lambda}{2}}-\cos^2\Theta}
		=\dfrac{-i\sqrt{1-\dfrac{M^4e^{2i\Lambda}/\cos^2\dfrac{\Lambda}{2} }{M^2e^{i\Lambda}-\cos^2\dfrac{\Lambda}{2}\cos^2\etamine\cos^2\Theta}}}{\dfrac{M^2 e^{i\Lambda}}{\cos^2\dfrac{\Lambda}{2}}-\cos^2\Theta} \,.
		\label{khdispcomprBsimple}
\end{eqnarray} 
We can analytically find purely imaginary solutions of that equation, which is the continuity of the ratio $\dfrac{\ktilde_1}{A_1}=\dfrac{\ktilde_2}{A_2}$.
First we observe that if the fluid characteristics are the same in the two parts and the velocities opposite, for purely imaginary $\omega$, i.e., real $\Lambda \in (0,\pi)$, the relations $\ktilde_2=\ktilde_1^*$ and $A_2=A_1^*$ hold.
This means that the dispersion relation is equivalent to the requirement the ratio $\dfrac{\ktilde_1}{A_1}$ to be real. 

We can write the numerator as $\ktilde_1=-|\ktilde|e^{-i \mu}$, with $0<\mu<\pi$ such that the amplitude vanishes at $x\to+\infty$. 
In order for the ratio $\dfrac{\ktilde_1}{A_1}$ to be real, $A_1 e^{i \mu}$ should be real. 
\\ Substituting the expressions of $\ktilde$ and $A$, the previous two relations mean that the imaginary parts of 
$\left[1-\dfrac{M^4e^{-2i\Lambda}/\cos^2\dfrac{\Lambda}{2} }{M^2e^{-i\Lambda}-\cos^2\dfrac{\Lambda}{2}\cos^2\etamine\cos^2\Theta}\right]e^{2i \mu}$ and $\left[\dfrac{M^2 e^{-i\Lambda}}{\cos^2\dfrac{\Lambda}{2}}-\cos^2\Theta\right]e^{i \mu}$ are zero. 

Thus we arrive at the two relations that give $\Lambda$ in parametric form with parameter $\mu$ 
\begin{adjustwidth}{-\extralength}{0cm}\begin{eqnarray} 
		\tan(2\mu)= 
		\dfrac{ \sin(2\Lambda) \cos^2\etamine\cos^2\Theta-2M^2\tan\dfrac{\Lambda}{2}}
		{1-\dfrac{M^2\cos\Lambda}{\cos^2\dfrac{\Lambda}{2}}+\left[\cos(2\Lambda)-\dfrac{1+\cos\Lambda}{M^2}\cos\Lambda\right]\cos^2\etamine\cos^2\Theta +\cos^4\dfrac{\Lambda}{2}\dfrac{\cos^4\etamine\cos^4\Theta}{M^4}} \,,
		\\ \cot\mu= \cot\Lambda-\dfrac{\cos^2\Theta}{2 M^2\tan\dfrac{\Lambda}{2}}
		\,. 
\end{eqnarray}\end{adjustwidth} 

Eliminating $\mu$ we find a single relation for the growth rate, the quartic polynomial equation~(\ref{khdispcomprBlambda}).

Note that the equation~(\ref{khdispcomprBlambda}) can also be derived 
by squaring equation~(\ref{khdispcomprBsimple}), since the substitution $\omega=ik_zV_0\tan\dfrac{\Lambda}{2}$ helps to exclude the trivial solutions of the resulting polynomial. This proves that indeed the solution is purely imaginary and verifies the above derivation, which has the advantage of connecting the solution with the correct sign of $\ktilde$ and its argument $\mu$.

\begin{adjustwidth}{-\extralength}{0cm}
\printendnotes[custom] 

\reftitle{References} 




 
\vspace{-1mm}
\PublishersNote{}
\end{adjustwidth}
\end{document}